\documentclass[aps,prl,amsmath,amssymb,twocolumn,preprintnumbers,nofootinbib]{revtex4}
\pdfoutput=1
\usepackage{graphicx}
\usepackage{multirow}
\usepackage{color}
\newcommand{\beq}{\begin{eqnarray}}
\newcommand{\eeq}{\end{eqnarray}}

\newcommand{\bmp}{\noindent\begin{minipage}{16cm}}
\newcommand{\emp}{\end{minipage}\vskip 7mm} % 7mm untightened

\usepackage{dcolumn}% Align table columns on decimal point
\usepackage{bm}% bold math
\usepackage{bbm}
\usepackage{pxfonts}
\usepackage[margin=5pt, font=normalsize,labelfont=bf,format=plain,justification=centerlast]{caption}
\usepackage{subfig}

\definecolor{rossoCP3}{cmyk}{0,.88,.77,.40}

\def\lsim{\mathrel{\rlap{\lower4pt\hbox{\hskip1pt$\sim$}}
    \raise1pt\hbox{$<$}}}                % less than or approx. symbol
\def\gsim{\mathrel{\rlap{\lower4pt\hbox{\hskip1pt$\sim$}}
    \raise1pt\hbox{$>$}}}                % greater than or approx. symbol

\newcommand{\be}{\begin{eqnarray}}
\newcommand{\ee}{\end{eqnarray}}

\usepackage[%dvips,
bookmarks]{hyperref}
\usepackage{color}
\hypersetup{colorlinks, bookmarksopen, bookmarksnumbered,
citecolor=verdes, linkcolor=bluc, pdfstartview=FitH, urlcolor=rossos}

\definecolor{grigio}{cmyk}{0,0,0,0.1}
\definecolor{rosa}{cmyk}{0,0.1,0.1,0.02}
\definecolor{rosino}{cmyk}{0,0.05,0.05,0.02}
\definecolor{rosas}{cmyk}{0,0.3,0.25,0.05}
\definecolor{celeste}{cmyk}{0.1,0,0,0.02}
\definecolor{giallino}{cmyk}{0,0,0.1,0.02}
\definecolor{rosso}{cmyk}{0,1,1,0.4}
\definecolor{rossos}{cmyk}{0,1,1,0.55}
\definecolor{rossoc}{cmyk}{0,1,1,0.2}
\definecolor{blu}{cmyk}{1,1,0,0.3}
\definecolor{blus}{cmyk}{1,1,0,0.5}
\definecolor{bluc}{cmyk}{1,1,0,0.1}
\definecolor{blucc}{cmyk}{0.7,0.5,0,0}
\definecolor{viola}{cmyk}{0,1,0,0.6}
\definecolor{viola2}{cmyk}{0,1,0.2,0.6}
\definecolor{verde}{cmyk}{0.92,0,0.59,0.25}
\definecolor{verdec}{cmyk}{0.92,0,0.59,0.15}
\definecolor{verdes}{cmyk}{0.92,0,0.59,0.4}
\definecolor{verdino}{cmyk}{0.12,0,0.09,0.02}
\definecolor{giallo}{cmyk}{0,0,1,0}
\definecolor{gialloverde}{cmyk}{0.44,0,0.74,0}
\definecolor{Titolo}{rgb}{0.752941176,0.576470588,0.992156863}% #C093FD
\definecolor{altro}{rgb}{0.094117647,0.650980392,0.643137255}% #24A6A4
\definecolor{Peanuts}{rgb}{0.2, 0.4, 0.6}% #336699
\definecolor{Pean1}{rgb}{0.6, 0.8, 0.4}% #99cc66
\definecolor{BHO}{rgb}{0.2, 0.8, 1}% #33CCFF
\definecolor{Daria}{rgb}{0, 0.9412, 0}% #00F000 e' uguale al verde u_u
\definecolor{UniPi}{rgb}{0.2549, 0.4627, 0.6275}% #4176a0
\definecolor{UniPidue}{rgb}{0.3216, 0.5804, 0.7882}% #5294c9

\newcommand{\ud}{\text{d}}

\usepackage{fancyhdr}
\begin{document}
%%%%%%%%%%%%%%%%%%%%%%%%%%%%%%%%%%%%%%%%%%%%%%%%%%%%%%%%%%%%%%%%%%%%%%%%%%%
\title{\Large  \color{rossoCP3}  Interfering Composite Asymmetric Dark Matter \\ for \\ DAMA and CoGeNT}
\author{Eugenio {\sc Del Nobile}}
%\email{delnobile@cp3-origins.net} 
\author{Chris {\sc Kouvaris}}
%\email{kouvaris@cp3-origins.net} 
\author{Francesco {\sc Sannino}}
%\email{sannino@cp3-origins.net} 
\affiliation{
{CP}$^{ \bf 3}${-Origins} and the Danish Institute for Advanced Study, University of Southern Denmark, Campusvej 55, DK-5230 Odense M, Denmark}
%%%%%%%%%%%%%%%%%%%%%%%%%%%%%%%%%%%%%%%%%%%%%%%%%%%%%%%%%%%%%%%%%%%%%%%%%%%%%%%%%%%%%%%%
\begin{abstract}
We provide a simple mechanism for reconciling the direct dark matter experimental results. We consider light asymmetric composite dark matter which scatters off nuclei via Higgs and photon exchange. We demonstrate that the interference between these two channels naturally accommodates the experimental results. We discover that this happens for a compositeness scale of the order of the electroweak. We also provide a model realization based on strong dynamics at the electroweak scale. 
\\[.1cm]
{\footnotesize  \it CP$^3$-Origins-2011-18  \&  DIAS-2011-04}
\end{abstract}

\maketitle
\thispagestyle{fancy}

%\section{Introduction}

The identification of dark matter (DM) is an important problem in modern physics. Apart from the key role DM plays in large structure formation and the evolution of the Universe, it might provide a link to
physics beyond the Standard Model (SM). Therefore it is not a surprise that so much experimental, observational, and theoretical effort has been devoted to the discovery of DM. There is strong 
evidence that DM might be in the form of Weakly Interacting Massive Particles (WIMPs), although other options are possible \cite{Visinelli:2009zm}. 

If DM is
indeed in the form of WIMPs, it can be characterized by the different properties it bears. It can be stable or decaying~\cite{Nardi:2008ix}.
It can also be produced thermally or non-thermally in the early Universe. Thermal production signifies that the relic abundance 
of DM is governed by annihilation between WIMPs. On the contrary, non-thermal production requires either a decay
of a heavier thermally produced particle, or the existence of an asymmetry between particles and antiparticles. Annihilations might kill the antiparticle population leaving out only particles to account for the DM density, thus the term asymmetric. This type of candidates appeared first in \cite{Nussinov:1985xr} in the form of technibaryons, and in \cite{Gudnason:2006ug} as Goldstone bosons. Since then, asymmetric DM candidates of all types have appeared in the literature ~\cite{Gudnason:2006ug, Foadi:2008qv,Khlopov:2008ty,Dietrich:2006cm,Sannino:2009za,Ryttov:2008xe,Kaplan:2009ag,Frandsen:2009mi,Belyaev:2010kp}.
We should note that the possibility of mixed DM with a thermally produced symmetric component and an asymmetric component \cite{Belyaev:2010kp}, or an asymmeric WIMP component and an asymmetric strongly interacting massive  
component ~\cite{Khlopov:2008ty} is also viable. 

From the experimental perspective, the situation is still unclear. Several experiments such as CDMS ~\cite{Ahmed:2010wy}, and Xenon10/100~\cite{Angle:2011th,Aprile:2011hi} find null evidence for DM, imposing
thus severe constraints on WIMP-nucleons cross sections, while DAMA~\cite{Bernabei:2008yi} and CoGeNT~\cite{Aalseth:2010vx} detect events that can be attributed to WIMP-nuclei collisions. DAMA has observed an annual modulation of the signal as it is expected due to the relative motion of the Earth with respect to the DM halo.  Recently CoGeNT confirmed the same modulation. Both experiments suggest a light WIMP of a mass of a few GeV. This makes asymmetric DM even more attractive since for such light WIMPs, a common mechanism for baryogenesis and DM production might take place. However, two comments are in order. The first is that the WIMP-nucleon cross sections required by DAMA and GoGeNT have been excluded by CDMS and Xenon upon assuming spin-independent interactions between WIMPs and nuclei (with protons and neutrons coupling  similarly to WIMPs). In that case the WIMP-nucleus cross section scales as $A^2$, where $A$ is the atomic number. Apart from non-WIMP scenarios that can explain this discrepancy~\cite{Khlopov:2010pq}, one proposed solution is that of inelastic DM~\cite{TuckerSmith:2001hy}, although this possibility has become recently more unlike. The second point is that DAMA and CoGeNT do not agree on the required WIMP-nucleon cross section upon assuming that protons and neutrons couple with equal strength to the WIMP. This is depicted in the top panel of Fig.~\ref{fig:experiments}. 

Generally speaking although such interactions where protons and neutrons are indistinguishable do exist, e.g. in the case of a Higgs exchange, other interactions can potentially distinguish protons from neutrons. Examples of the latter case are the photon exchange (that obviously couples only to the protons, and therefore the cross section will scale as $Z^2$, where $Z$ is the number of protons), and the $Z$-boson exchange that couples protons and neutrons differently, having a scaling for the cross section as  $(A-Z+ \epsilon Z)^2$ where $\epsilon = 1-4 \sin^2 \theta_W \sim 0.08$ ($\theta_W$ being the Weinberg angle). However, due to the fact that in most stable nuclei the number of protons is close to the one of neutrons, the discrepancy between DAMA/CoGeNT and CDMS/Xenon remains when considering the cross section from each interaction separately. 

Recently, it was observed in \cite{Chang:2010yk,Feng:2011vu} that a relative strength of the couplings of protons and neutrons $f_n/f_p\simeq -0.7$ can cause an overlap of the DAMA and GoGeNT regions. {In \cite{Frandsen:2011ts} it was argued that inelastic DM can enhance the annual modulation fraction bringing CoGeNT an CDMS results into a better agreement.}

In this paper we present a quite generic and well motivated model of composite light asymmetric DM where we achieve an overlap of the DAMA and GoGeNT regions via interference between two channels of interactions. Our DM candidate although electrically neutral, it carries electric dipole moment and can exchange a photon with the protons of the nucleus it scatters off. In addition, the particle couples also to the Higgs boson. We shall demonstrate on quite general grounds that for a composite scale of the order of the electroweak, interference between Higgs and photon exchange can naturally explain the $f_n/f_p\simeq -0.7$ fitting value, evading simultaneously the CDMS and Xenon constraints.  We also provide a natural model where such a light asymmetric DM candidate emerge. Due to the fact that our WIMP will be a composite particle made of fermions with spin-independent interactions, recent astrophysical constraints on asymmetric DM from observations of neutron stars~\cite{Kouvaris:2010jy} are avoided.
\vspace{ -.3cm}
 \section{Interfering Dark Matter}
\vspace{-.2cm}
  On general grounds, one can write an effective theory for a composite scalar DM (with composite scale $\Lambda$) interacting with the SM fields and with itself. The first comprehensive effective theory appeared in \cite{DelNobile:2011uf}. 
 According to this classification the relevant operators for scattering on nuclei involve the coupling with the Higgs boson and the photon. The interaction with the photon occurs via a dimension 6 dipole-type term. This term appears naturally in any model of composite DM similar to Technicolor Interacting Massive Particles (TIMP)~\cite{Foadi:2008qv}.  
We assume that the DM scalar  $\phi$ is neutral under the SM but charged under an extra $U(1)_{\phi}$ global symmetry protecting it against decay. The leading terms relevant for this analysis are: 
% \begin{widetext}
 \begin{equation}\label{Lagrangian}
\mathcal{L} = \partial_\mu \phi^* \partial^\mu \phi - (M_\phi^2 - d_H \frac{v_{EW}^2}{2}) \phi^* \phi - d_H H^\dagger H \phi^* \phi + \frac{d_B}{\Lambda^2} e J_\mu \partial_\nu F^{\mu\nu} \ ,
\end{equation}
%\end{widetext}
where $H = (\pi^+, \frac{1}{\sqrt{2}} (v_{EW} + h + i \, \pi^0))$ is the Higgs doublet, $J_\mu = i \, \phi^* \overleftrightarrow{\partial_\mu} \phi$, $F^{\mu\nu}$ is the photon field strength, and $v_{EW} \simeq 246$~GeV. 

The zero momentum transfer cross section of a WIMP scattering off a nucleus with $Z$ protons and $A - Z$ neutrons is \cite{Foadi:2008qv, Belyaev:2010kp}
\begin{equation}\label{sigma_A}
\sigma_A = \frac{\mu_A^2}{4 \pi} \left| Z f_p + (A - Z) f_n \right|^2 \ ,
\end{equation}
where
\beq
f_n = d_H f \frac{m_p}{m_H^2 M_\phi} \ , \qquad\qquad f_p = f_n - \frac{8 \pi \alpha d_B}{\Lambda^2} \ , \label{fnfp}
\eeq
$m_p$ is the nucleon mass, $\mu_A$ is the WIMP-nucleus reduced mass and~$f \sim 0.3$~parametrizes the Higgs to nucleon coupling. 

The event rate for generic couplings $f_n$ and $f_p$ is 
\beq\label{R}
R = \sigma_p \sum_i \eta_i \frac{\mu_{A_i}^2}{\mu_p^2} I_{A_i} \left| Z + (A_i - Z) f_n / f_p \right|^2 \ ,
\eeq
where $\eta_i$ is the abundance of the specific isotope $A_i$ in the detector material, and $I_{A_i}$ contains all the astrophysical factors as well as  the nucleon form factor $ F_{A_i} (E_R)$. For a given isotope we have
\beq
I_{A_i} = N_T \, n_\phi \int \ud E_R \int_{v_\text{min}}^{v_\text{esc}} \ud^3 v \, f(v) \frac{m_{A_i}}{2 v \mu_{A_i}^2} F_{A_i}^2 (E_R) \ .
\eeq
Here $m_{A_i}$ is the mass of the target nucleus, $N_T$ is the number of target nuclei, $n_\phi$ is the local number density of DM particles, and $f(v)$ is their local velocity distribution. The velocity integration is limited between the minimum velocity required in order to transfer a recoil energy $E_R$ to the scattered nucleus, $v_\text{min} = \sqrt{m_A E_\text{R} / 2 \mu_A^2}$, 
and the escape velocity from the galaxy $v_\text{esc}$. The WIMP-proton cross section $\sigma_p =  {\mu_p^2} \left| f_p \right|^2 /{4 \pi}$ can be easily obtained by setting $A = Z = 1$ in Eq.~\eqref{sigma_A}.
 
\begin{figure}
\begin{center}
\includegraphics[width=.35\textwidth%, height=0.7\textwidth
]{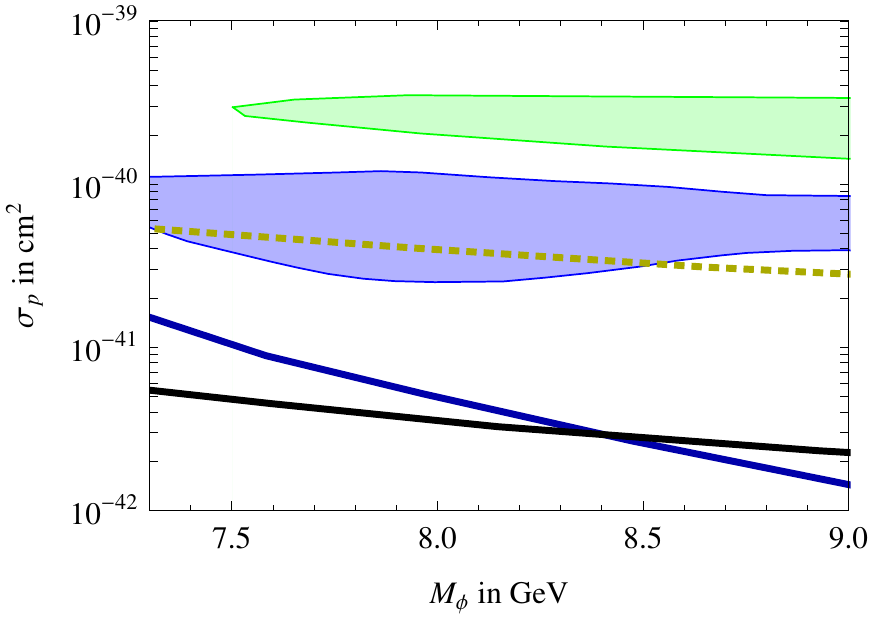}
%\bigskip
\includegraphics[width=.35\textwidth%, height=0.7\textwidth
]{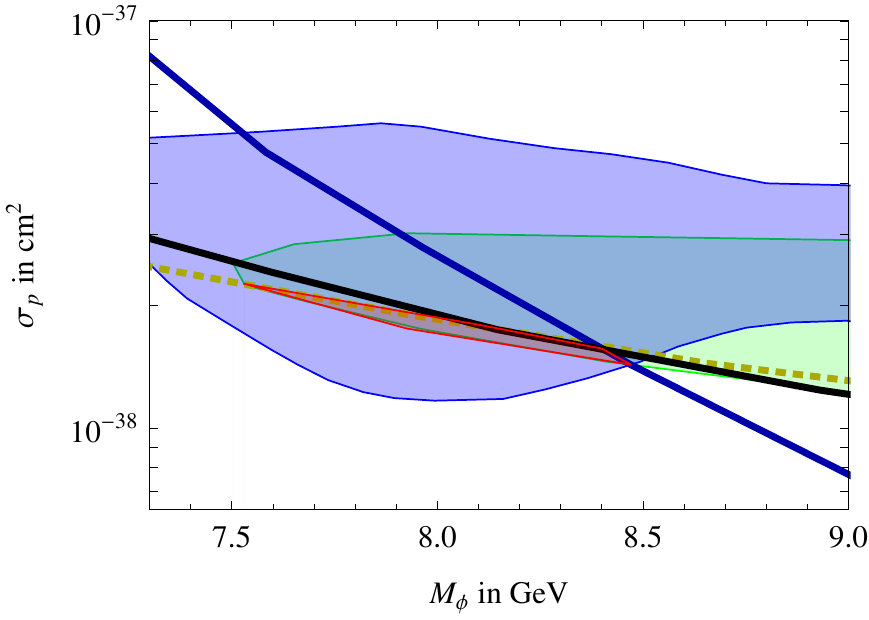}
\caption{\em\label{fig:experiments}Favored regions and exclusion contours in the $(M_\phi, \sigma_p)$ plane for the standard case $f_n / f_p = 1$ (top panel) and the case $f_n / f_p = - 0.71$ (bottom panel). The green contour is the $3 \sigma$ favored region by DAMA/LIBRA \cite{Savage:2010tg} assuming no channeling \cite{Bozorgnia:2010xy} and that the signal arises entirely from Na scattering; the blue region is the $90 \%$ CL favored region by CoGeNT; the dashed line is the exclusion plot by CDMS II Soudan \cite{Ahmed:2010wy}; and the black and blue lines are respectively the exclusion plots from the Xenon10 \cite{Angle:2011th} and Xenon100 \cite{Aprile:2011hi} experiments. The common region passing all the constraints is shown in red.}
\end{center}
\end{figure}

Direct DM search collaborations quote constraints on WIMP-nuclei cross sections normalized to the WIMP-nucleon cross section $\sigma_p^\text{exp}$ (assuming conventionally  $f_n = f_p$).  Therefore the experimentally constrained event rate can be casted in the following form
\beq \label{Rexp}
R = \sigma_p^\text{exp} \sum_i \eta_i \frac{\mu_{A_i}^2}{\mu_p^2} I_{A_i} A_i^2 \ .
\eeq
  Equating Eqs.~\eqref{R} and \eqref{Rexp} yields the experimental constraints on the generic WIMP-proton cross section $\sigma_p$ (with arbitrary couplings $f_p$ and $f_n$)
\beq\label{sigma_pTOT}
\sigma_p = \sigma_p^\text{exp} \frac{\sum_i \eta_i \mu_{A_i}^2 I_{A_i} A_i^2}{\sum_i \eta_i \mu_{A_i}^2 I_{A_i} \left| Z + (A_i - Z) f_n / f_p \right|^2} \ .
\eeq
Provided that the factors $I_{A_i}$ do not change significantly from one  isotope to another (as we checked), they cancel out from numerator and denominator.
In the top panel of Fig.~\ref{fig:experiments} we plot the exclusion limits from CDMS II and Xenon10/100, and the favored regions of DAMA and CoGeNT in the $(M_{\phi},\sigma_p)$ plane for $f_n/f_p=1$. The DAMA and CoGeNT regions do not coincide. However as it was first suggested in \cite{Chang:2010yk,Feng:2011vu} possible variation of $f_n/f_p$ can move the two regions around. We confirm that for $f_n/f_p=-0.71$, the DAMA/LIBRA and CoGeNT regions partially overlap, leaving even a small region of phase space that evades  the tightest bounds coming from CDMS II and Xenon10/100.
 The isotopic abundances $\eta_i$ we use are provided in \cite{Feng:2011vu}.
\\
In the small allowed region of the phase space for the optimal value $f_n / f_p=-0.71$,  the WIMP mass $M_\phi$ ranges between $7.5$ and $8.5$ GeV, and  the WIMP-proton cross section $\sigma_p$ is $\sim 2 \times 10^{-38}$ cm$^2$. There is not much freedom to change $f_n / f_p$, since even small changes in the ratio drive the DAMA/CoGeNT overlapping region within the excluded area by either CDMS or Xenon.  For example, for $f_n/f_p=- 0.70$ CDMS II excludes the whole DAMA region, while for $f_n/f_p=- 0.72$ the Xenon10 line excludes both DAMA/LIBRA and CoGeNT.
Fixing the ratio $f_n / f_p=-k $ leads to a constraint on the parameters of Eq.~(\ref{fnfp})
\beq
d_H = \frac{8 \pi \alpha \, k \, m_H^2 M_\phi}{f (1 + k) m_p} \frac{d_B}{\Lambda^2} \ .
\eeq
Fixing $\sigma_p \sim 2 \times 10^{-38}$ cm$^2$ provides an extra condition which allows us to determine both $d_B$ and $d_H$, so long we fix the Higgs mass and the scale $\Lambda$
\beq
\frac{d_B}{\Lambda^2} \sim 3 \times 10^{-4} \text{ GeV}^{-2} \ ,  ~~~  \frac{d_H}{m_H^2} \sim 5 \times 10^{-4} \text{ GeV}^{-2} \ .
\eeq
Assuming a Higgs mass of the order of  $\mathcal{O}(100 \text{ GeV})$ and $\Lambda \sim v_{EW}$ we find $d_B \sim d_H \sim \mathcal{O}(1) - \mathcal{O}(10)$. Therefore  interfering DM emerging from composite dynamics at the electroweak scale can resolve the experimental puzzle.  
\vspace{-.3cm}
 \section{A Model for DAMA and CoGeNT}
\vspace{-.2cm}
Light DM particles are natural in several extensions of the SM. One particularly appealing possibility is a candidate emerging from new strong dynamics at the electroweak scale such as Technicolor (see \cite{Sannino:2009za} for a recent review). Here we use as specific model Ultra Minimal Technicolor (UMT) \cite{Ryttov:2008xe}. 
%The model constitutes an explicit example of (near) conformal strong dynamics featuring two types of technifermions, i.e. transforming according to two different representations of the underlying technicolor gauge group \cite{Dietrich:2006cm}. 
The model
\begin{itemize}
\item features the lowest value of the $S$ parameter in Technicolor theories \cite{Sannino:2010ca};
 
% \item contains the minimal number of fermions needed to achieve conformal dynamics;

\item yields several natural DM candidates of either symmetric, antisymmetric or mixed type \cite{Belyaev:2010kp};

\item allows for multiple electroweak finite temperature phase transitions \cite{Jarvinen:2009pk}. 

\end{itemize}
  A later variation  of this model appeared in \cite{Galloway:2010bp} under the name of Minimal Conformal Technicolor.  UMT  is constituted by an $SU(2)$ technicolor gauge group with two Dirac flavors in the fundamental representation also carrying electroweak charges, as well as, two additional Weyl fermions in the adjoint representation but singlets under the SM gauge group. The overall global symmetry is $SU(4)\times SU(2) \times U(1)$ which breaks spontaneously to $Sp(4)\times SO(2)\times Z_2$. We focus here on the $SU(4)$ to $Sp(4)$ sector which is the one responsible for electroweak symmetry breaking. Five Goldstone bosons are generated and three become the longitudinal degrees of freedom of the SM gauge bosons. The remaining two Goldstone bosons are arranged in a complex scalar, which we identify with the light DM candidate $\phi$, carrying the $U(1)_{\phi}$ global symmetry included in the original $SU(4)$ flavor symmetry. The $U(1)_{\phi}$ global symmetry corresponds to the technibaryon number associated to $\phi$ which, in terms of the underlying technifermions, is a di-techniquark. A mass term for $\phi$ comes from the new sector responsible for giving masses to the SM fermions and from electroweak corrections. Being a pseudo Goldstone boson it has therefore either derivative interactions, or non derivative ones with couplings vanishing with $M_{\phi}$.  The effective Lagrangian is \cite{Belyaev:2010kp}
  \begin{eqnarray}
  {\cal L} & = & \partial_{\mu}\phi^{\ast} \partial^{\mu}\phi - M^2_{\phi}
\phi^{\ast} \phi + \frac{d_1}{\Lambda}  h \, \partial_{\mu}\phi^{\ast}\partial^{\mu}\phi  \\
&&+  \frac{d_2}{\Lambda} M^2_{\phi}\, h \,\phi^{\ast}\phi  
+  \frac{d_3}{2\Lambda^2} h^2 \partial_{\mu}\phi^{\ast} \partial^{\mu}\phi  
+  \frac{d_4}{2\Lambda^2} M^2_{\phi}h^2 \phi^{\ast} \phi  \ . \nonumber 
\end{eqnarray}
This Lagrangian provides the correct number of independent operators. In the nonrelativistic limit the Lagrangian \eqref{Lagrangian} and the one above give the same WIMP-nucleus cross section provided we set
   \begin{eqnarray}
  d_H  &=&  -\frac{d_1 + d_2}{v_{EW}\,\Lambda} M^2_{\phi}
  %  = \frac{d_3 + d_4}{\Lambda^2} M^2_{\phi} 
  \ .
  \end{eqnarray}
  
 Assuming that the physics is such that in the early universe at temperatures higher than the electroweak symmetry breaking scale there has been a mechanism which has led to an asymmetry in either the baryon, the $\phi$, or the lepton number, electroweak sphaleron processes will have equilibrated the different numbers. According to the estimates of \cite{Ryttov:2008xe} we have
\begin{equation}
\frac{\Phi}{B} = \frac{\sigma}{2} \left(3+\frac{L}{B}\right) \ ,
\end{equation}
with $\Phi$ indicating the technibaryon number, $L$ the lepton number, $B$ the baryonic one and $\sigma$ the statistical function of the techniquarks depending on the ratio between the techniquark dynamically generated mass and the temperature below which the electroweak sphaleron processes cease to be relevant.  The ratio of dark to baryon energy density is then
\begin{equation}
\frac{\Omega_{DM}}{\Omega_{B}} = \frac{\Omega_{\phi}}{\Omega_{B}} = \frac{M_{\phi}}{m_p}\frac{\Phi}{B} \ .
\end{equation}
For techniquark masses of the order of the electroweak scale and assuming zero lepton number 
 we have
$
%\begin{equation}
M_{\phi} \sim 3.3 \, m_p
%\end{equation}
$ for $\Omega_{DM} \simeq  5 \Omega_{B}$. The results are valid for either a second order or a first order electroweak phase transition \cite{Ryttov:2008xe}.  It is easy to check that with $L \sim -2 B$ one gets $M_\phi \sim 8$~GeV.  Here we are assuming that the annihilation cross section is sufficiently large to eliminate any symmetric component. This can be achieved either by reducing the composite Higgs mass or by increasing the size of the coupling to the SM fields \cite{Belyaev:2010kp}. Collider signatures of this type of DM have been studied in \cite{Frandsen:2009mi,Foadi:2008qv}. 

% Save this file and include it in your paper as the bibliography
% or cut and paste directly into your LaTeX

\end{document}